\documentclass[12pt,fleqn]{article}
\usepackage{amsmath,amssymb,amsthm,epsfig,graphics}
\usepackage{lscape}

\newcommand{\ri}{{\mathrm i}}
\newcommand{\p}{\partial}
\newcommand{\bea}{\begin{array}}
\newcommand{\eea}{\end{array}}

\catcode`@=11 \long
\def\@caption#1[#2]#3{\par\addcontentsline{\csname
ext@#1\endcsname}{#1} {\protect\numberline{\csname
the#1\endcsname}{\ignorespaces #2}} \begingroup \small
\@parboxrestore \@makecaption{\csname fnum@#1\endcsname}
{\ignorespaces #3}\par \endgroup} \catcode`@=12

\newcommand{\la}{\label}

\catcode`@=11 \long
\def\@caption#1[#2]#3{\par\addcontentsline{\csname
ext@#1\endcsname}{#1} {\protect\numberline{\csname
the#1\endcsname}{\ignorespaces #2}} \begingroup \small
\@parboxrestore \@makecaption{\csname fnum@#1\endcsname}
{\ignorespaces #3}\par \endgroup} \catcode`@=12

\begin{document}

\allowdisplaybreaks
 \begin{titlepage} \vskip 2cm

\begin{center} {\Large\bf The maximal 'kinematical' invariance group for
an arbitrary potential revised}
\footnote{E-mail: {\tt nikitin@imath.kiev.ua} } \vskip 3cm {\bf {A.
G. Nikitin } \vskip 5pt {\sl Institute of Mathematics, National
Academy of Sciences of Ukraine,\\ 3 Tereshchenkivs'ka Street,
Kyiv-4, Ukraine, 01601\\}}\end{center} \vskip .5cm \rm

\begin{abstract}
Group classification of one particle Schr\"odinger equations with arbitrary potentials (C. P. Boyer,  Helv. Phys. Acta {\bf 47}, p. 450, 1974) is revised. The corrected completed list of non-equivalent potentials and the corresponding symmetries is presented together with exact identification of symmetry algebras and admissible equivalence transformations.
\end{abstract}
\end{titlepage}
\section{Introduction\label{int}}
One of key concepts of any consistent physical theory is symmetry. For example, relativistic theories should possess  Lorentz invariance, which is replaced by Galilei invariance for systems with velocities much less than the velocity of light.

A regular way for searching of continuous and some other symmetries was proposed long time ago by Sophus Lie. In particular, he created the grounds of the group classification of differential equations. Being applied to model equations of mathematical and theoretical physics, it presents an effective way for construction of theories with a priori requested symmetries.

A perfect example of a group classification of fundamental equations of physics was the completed description of symmetries which can be possessed  by  Schr\"odinger equations  with arbitrary potentials. It has been done more than forty  years ago
in papers \cite{Nied}, \cite{And} and
\cite{Boy}. It was Niederer \cite{Nied} who had found the maximal invariance group of the free Schr\"odinger equation. He  was the first who demonstrated that this group is more extended  than just the Galilei group discussed previously in \cite{LL}, and includes also dilations and conformal transformations. Notice that in fact this result was predicted by Sophus Lie who described symmetries of the heat equation.

Symmetries of Schr\"odinger equation with non-trivial potential were described in paper \cite{And}, but only for the case of one spatial variable. Then Boyer extends these results for  more physically interesting planar and three dimensional systems.

The mentioned contributions had a big impact and occupy a place of honour in modern physics. They present a priori information about possible symmetries of  one particle QM systems and so form group-theoretical grounds of quantum mechanics. The group classification is the necessary step in investigation of  higher symmetries
of Schr\"odinger equations requested for separation of variables \cite{Mil} and description of superintegrable systems \cite{Wint}. They also give rise for some new inspiring physical theories such
as the Galilean conformal field theory \cite{Hag}.

Recently we extend the Boyer classification  to the case of QM systems with position dependent mass \cite{NZ}, \cite{NZ2}, \cite{N3}. It was a good opportunity to revise the classical paper \cite{Boy} bearing in mind  that some results of this paper appear as particular cases of our analysis. As a result it was recognized, that the Boyer classification is incomplete and some systems with inequivalent symmetries are missing in \cite{Boy}. In addition, the classification results are presented in a rather inconvenient form. In the classification table there is a list of potentials while the symmetry generators are absent, and the reader is supposed to look for them in different places of the paper text an make rather nontrivial speculation to identify them. Moreover, this identification is not well-defined, and some potentials include parameter which can be removed using equivalence transformations.

We believe that the physical community can pretend to a conveniently presented and correct information on point continuous symmetries which can be admitted by the main equation of quantum mechanics. It was the main reason to write the present work, where the Boyer results are verified and corrected. In addition, the methods of group analysis of differential equations are essentially developed in comparison with the seventieth  of the previous century, and it was interesting to apply them to the well known and very important object of mathematical physics.
\section{Determining equations}
We will  consider Schr\"odinger equations of the following generic form
\begin{gather}\left(\ri\frac{\p}{\p t}-H\right)\psi(t,{\bf x})=0\la{se}\end{gather}
where $H$ is the Hamiltonian 
\begin{gather}H=-\frac12 \p_a\p_a
+V({\bf x})\la{H}\end{gather}
with
\begin{gather*}\p_a=\frac{\p}{\p{x_a}},\quad {\bf x}=(x_1, x_2,...,x_n)\end{gather*}
and summation is imposed over the repeating indices $a$ over the values $a=1, 2,...,n$.

We will search for symmetries of equations (\ref{se}) with respect
to continuous groups of transformations of dependent and independent variables. We will not apply the generic Lie approach but, following \cite{Boy}, restrict ourselves to using its simplified version which is perfectly adopted to equation (\ref{se}). Let us represent the infinitesimal operator of the searched transformation group in the form
\begin{gather}\label{so}
    Q=\xi^0\partial _t+\xi^a\partial_a+\tilde\eta\equiv \xi^0\partial _t+
    \frac12\left(\xi^a\partial_a+\partial_a\xi^a\right)+\ri\eta,
\end{gather}
where $\tilde
\eta=\frac12\xi^a_a+\ri\eta,\ \ $ $\xi^0$, $\xi^a$ and $\eta$ are
functions of independent variables and $\p_t=\frac{\p}{\p t}$.

Generator (\ref{so}) transforms solutions of equation (\ref{se}) into solutions if it satisfies the  following operator equation
\begin{gather}\la{ic}QL-LQ=\alpha L\end{gather}
where $L=\ri\p_t-H$ and $\alpha$ is one more unknown function of $t$ and $\bf x$.

Evaluating the commutator in the l.h.s. of (\ref{ic}) and equating coefficients for the linearly independent
differentials we obtain the following system of  equations for unknowns $\xi^0, \xi^a, \eta,\  V $ and $\alpha$:
\begin{gather}
\dot\xi^0=-a, \quad  \xi^0_a=0,\la{de1}\\
\la{de2} \xi^b_{a}+\xi^a_{b}-\frac{2}n\delta_{ab}\xi^i_i=0,
\\\label{de6} \xi^i_i=-\frac{n}2\alpha,\\ \dot\xi^a+\eta_a
=0,\label{de7}\\
\label{de8}\xi^aV_a=\alpha V+\dot\eta.\end{gather}

 In accordance with (\ref{de1}) both $\xi^0$ and $\alpha$ do not depend on $\bf x$. Equations (\ref{de2}) and (\ref{de6}) specify the $x$--dependence of coefficients $\xi^a$:
\begin{gather}\la{kil} \xi^a=
-\frac\alpha2 x_a+\theta^{ab} x_b+\nu_a\end{gather}
where $\alpha$, $\theta^{ab}=-\theta^{ba}$ and $\nu^a$ are (in general time dependent) parameters. Moreover, in accordance with (\ref{de7}), $\theta^{ab}$ are time independent, and so we will deal with five unknown functions $\alpha, \eta$  and $\nu^a$ depending on $t$. Then, integrating  (\ref{de7}) we obtain the generic form of function $\eta$:
\begin{gather}\la{eta}\eta=\frac{\dot\alpha}4x^2-\dot\nu_ax_a+f(t),\end{gather}
and equation (\ref{de8}) is reduced to the following form:
\begin{gather}\la{ded}\left(\frac\alpha2x_a-\nu_a-\theta^{ab}x_b\right)V_a+\alpha V+\frac{\ddot \alpha}4x^2-\ddot\nu_ax_a-\dot f=0.\end{gather}

Thus to classify point Lie symmetries of equation (\ref{se}) it is necessary to find all non-equivalent solutions of equation (\ref{ded}). The evident equivalence transformations for (\ref{se}), i.e., transformations which keep the generic form of this equation but can  change the potential, are shifts, rotations and scalings of independent variables. Such transformations of the spatial variables form the Euclid group E(n) extended by simultaneous  scaling of these  variables.  In addition, we can scale $\psi$ and make the following transformations:
 \begin{gather}\la{shift}\psi\to\exp(\ri t C)\psi,\quad  V\to V+C\end{gather} where $C$ is a constant.

 It is not difficult to show that the mentioned transformations exhaust the equivalence group for equation (\ref{se}) with arbitrary potential. However, for some particular potentials there are additional  equivalence transformations which will be specified  in the following.

\section{Symmetries for equations with trivial and isotropic oscillator potentials}
Setting in  (\ref{ded}) $V=0$ we can easy solve the obtained equation and find the corresponding admissible symmetries (\ref{so}). They are linear combinations of the following symmetry operators:
\begin{gather}\begin{split}& P_a=-\ri\p_a,\\&M_{ab}=x_aP_b-x_bP_a,\\ &D=2tP_0-x_aP_a+\frac{\ri n}2, \end{split}\label{sok}\\\begin{split}
 &P_0=\ri\p_t,\quad G_a=tP_a-x_a,\\&
A=t^2P_0-tD+\frac{x^2}2.\end{split}\la{sos}\end{gather}

Operators (\ref{sok}), (\ref{sos}) together with the unit operator $I$ satisfy the following commutation relations:
\begin{gather}\la{cr1}\quad [P_0,A]=\ri D,\quad [P_0,D]=2\ri P_0, \quad [D,A]=2\ri A,\\\vspace{1mm}\la{cr2}
 \quad[P_a,D]=\ri P_a,\quad [G_a,D]=-\ri G_a,\\
 \begin{split}\la{cr3}&[P_0,G_a]=\ri P_a,\quad [P_a,G_b]=\ri\delta_{ab}I,\\
&[M_{ab},M_{cd}]=\ri(\delta_{ad}M_{bc}+\delta_{bc}M_{ad}
-\delta_{ac}M_{bd}-\delta_{bd}M_{ac}
 \\&[P_a,M_{bc}]=\ri(\delta_{ab}P_c-\delta_{ac}P_b),
 \quad [G_a,M_{bc}]=\ri(\delta_{ab}G_c-\delta_{ac}G_b),
 \end{split}\end{gather}
 (the remaining commutators are equal to zero) and
 form the Lie algebra schr(1,n) whose dimension is $N=\frac{n^2+3n+8}2$. Commutation relations (\ref{cr3}) specify the Lie algebra g(1,n) of Galilei group, which is a subalgebra of schr(1,n).

       Let us present additional identities satisfied by
 operators (\ref{sok}) and (\ref{sos}) (see, e.g., \cite{Nunu}):
\begin{gather}\la{ide1}P_aG_b-P_bG_a=M_{ab}, \\P_aG_a+G_aP_a=2D+2t(P^2-2P_0),\la{ide2}\\
G_aG_a=2A+t^2(P^2-2P_0)\la{ide3}.\end{gather}

On the set of solutions of equation (\ref{H}) the term in brackets is equal to $-2V$. Since in our case $V=0$  relations (\ref{ide1})--(\ref{ide3}) express  generators $M_{ab},\ D$ and $A$ via $P_a$ and $G_a$.

 Operators (\ref{sok}) and (\ref{sos}) generate the  $N-$parametric symmetry group which is much more extended than the equivalence group for equation (\ref{se}) with arbitrary potential. Moreover, there are three  additional equivalence transformations
\begin{gather}\la{et1}\begin{split}&{\bf x}\to \tilde {\bf x}=
\frac{\bf x}{\sqrt{1+t^2}}, \quad t\to\tilde
t=\frac1{\omega}{\arctan(
t)},\\&\psi(t,{\bf x})\to\tilde\psi(\tilde t,\tilde {\bf x})=(1+t^2)^{\frac{n}4}\text{e}^{\frac{-\ri\omega t{\bf x}^{2}}{2(1+
t^2)}}\psi(t,{\bf x}),
\end{split}\end{gather}
\begin{gather}\la{et2}\begin{split}&{\bf x}\to \tilde {\bf x}=
\frac{\bf x}{\sqrt{1-t^2}}, \quad t\to\tilde
t=\frac1{\omega}{\text{arctanh}(
t)},\\&\psi(t,{\bf x})\to\tilde\psi(\tilde t,\tilde {\bf x})=(1-t^2)^{\frac{n}4}\text{e}^{\frac{\ri\omega t{\bf x}^{2}}{2(1-
t^2)}}\psi(t,{\bf x})
\end{split}\end{gather}
and
\begin{gather}\la{et3}\begin{split}& x_a\to x'_a=x_a-\frac12{\kappa_a}t^2,\quad  t\to t'=t,\\& \psi(t,{\bf x})\to\psi'(t',{\bf x}')=\exp\left(-it\kappa_ax_a+\frac\ri3\kappa^2t^3\right)\psi(t,{\bf x})\end{split}\end{gather}

which keep the generic form of equation (\ref{se}) but change the trivial potential $V=0$ to
\begin{gather}\la{quad}V=\frac12\omega^2x^2,\\\la{qua}V=-\frac12\omega^2x^2,
\end{gather}
and
\begin{gather}\la{lin}V=\kappa_ax_a\end{gather}  correspondingly.

The transformations  which reduce the isotropic  harmonic and repulsive oscillators to the free particle Schr\"odinger equation  were discovered by Niederer \cite{Nied2}. Formulae (\ref{et1}) and (\ref{et2}) present transformations for wave functions dependent on $n$ spatial variables while in \cite{Nied2} we can find them only for $n=1$.
Let us note that in fact the origin of this transformation is much more extended: it can be applied for any  equation (\ref{se}) with potential $V$ being a homogeneous function   of degree -2.

Mapping (\ref{et3}) connects the systems with trivial and free fall potentials \cite{Nied3}. However, its origin can be extended to all potentials dependent on a reduced number $m$ of spatial variables with $m<n$.

Symmetries of equation (\ref{H}) with quadratic potential (\ref{quad}) can be obtained from (\ref{cr1})--(\ref{cr3}) applying transformations (\ref{et1}). They include $P_0$, $M_{ab}$ and generators  presented in the following formulae:
\begin{gather}\la{sos2}\begin{split}&A_{1}^{+}=\frac1{\omega}\sin(2\omega t)P_0-\cos(2\omega t)\left(x^aP_a-\frac{\ri n}2\right)-{\omega}\sin(2\omega t)x^2,\\
&A_{2}^{+}=\frac1{\omega}\cos(2\omega t)P_0+\sin(2\omega t)\left(x_aP_a-\frac{\ri n}2\right)-{\omega}\cos(2\omega t)x^2,\\
&B^+_{a}(\omega)=\sin(\omega t)P_a-\omega x_a\cos(\omega t),\quad \hat B^+_{a}(\omega)=\cos(\omega t)P_a+\omega x_a\sin(\omega t)\end{split}\end{gather}

For the case of the repulsive oscillator potential (\ref{qua}) we have the following  symmetries
\begin{gather}\la{sos3} <P_0,\ M_{ab},\ A_{1}^{-},\ A_{2}^{-},\ B^-_{a}, \ \hat B^-_{a}>\end{gather}
whose explicit form can be obtained from (\ref{sos2}) changing $\omega\to\ri\omega$.

Analogously, staring with  realization (\ref{sok}), (\ref{sos}) and making transformations (\ref{et3}) it is not difficult to find symmetries for equation (\ref{H}) with the free fall potential. We will not write the related  cumbersome expression which can be easily obtained making the changes
\begin{gather}\la{chacha}\begin{split}&P_0\to P_0 +\kappa_aG_a+\frac12\kappa^2t^2,\quad P_a\to P_a+\kappa_at,\quad x_a\to x_a+\frac12\kappa_a t^2\end{split}\end{gather}
in all generators (\ref{sok}) and (\ref{sos}).

\section{Classification results for arbitrary potentials}

Let us consider equation (\ref{ded}) for arbitrary potential $V$. Its terms are products of functions of different independent variables which makes it possible to make the effective separation of these variables and reduce the problem to solution of systems of ordinary differential equations for time dependent functions $\alpha,\ \nu^a$ and $f$. Then the corresponding potentials are easily calculated integrating equations (\ref{ded}) with found functions of $t$.

There are different ways to realize this programm. We use the algebraic approach whose main idea is to exploit the basic property of symmetry operators, i.e.,  the fact that they should form a basis of a Lie algebra. This algebra by definition includes operator $P_0$ and the unit operator.

In this section we preset all non-equivalent $3d$ and $2d$ potentials which correspond to more extended symmetries. They are collected in the following Tables 1 and 2, where $G(.)$ are arbitrary function of variables given in the brackets, $\mu, \ \kappa $ and $\omega_a$ are arbitrary real parameters, $\varepsilon_1,\ \varepsilon_2$ and $\varepsilon_3$ can take values $\pm 1$ independently, subindexes $a$ and $k$ take all values 1, 2, 3 and 1, 2 correspondingly. In addition, we denote $r=\sqrt{x_1^2+x_2^2+x_3^2}, \ \tilde r=\sqrt{x_1^2+x_2^2}$ and $\varphi=\arctan\left({x_2}/{x_1}\right)$.

\newpage
\begin{center}Table 1.
Non-equivalent potentials and  symmetries for 3$d$ Schr\"odinger equation.
\end{center}
\begin{tabular}{l l l l }
\hline No&Potential $V$&Symmetries&Invariance algebras
\\
\hline

1$\hspace{2mm} $&$G(\tilde r,x_3)+\kappa\varphi$
\vspace{1mm}
&$ L_3+\kappa t
$&\hspace{-2.5mm}$\begin{array}{c}\textsf{n}_{3,1}\ \ \text{if}\ \ \kappa\neq0,\\
3\textsf{n}_{1,1}\ \text{if}\ \ \kappa=0\end{array}$\\
2$\hspace{2mm} $&$G(\tilde r,x_3-
\varphi)+\kappa \varphi$
\vspace{1mm}
&$L_3+P_3+\kappa t$&$\hspace{-2.5mm}\begin{array}{c}\textsf{n}_{3,1}\ \ \text{if}\ \ \kappa\neq0,\\
3\textsf{n}_{1,1}\ \text{if}\ \ \kappa=0\end{array}$\\
3$\hspace{2mm} $&$\frac1{r^2} G(\frac{r}{\tilde r},r^\kappa e^
{-\varphi})$
 \vspace{2mm}
 &$D+\kappa L_3$&$\textsf{s}_{2,1}\oplus\textsf{n}_{1,1}$\\
 4$^\star\ $&$G(x_1,x_2)$
\vspace{1mm}
&$G_3,\ P_3$& $\textsf{n}_{4,1}$\\
5$^\ast\ $&$\frac1{ r^2}G(\varphi,\frac{\tilde r}r)$
\vspace{2mm}
&$A,\ D$&sl(2,R)$\oplus\textsf{n}_{1,1}$
\\
6$^\ast\ $&$\frac1{ r^2}G(\frac{\tilde r}r)$
\vspace{2mm}
&$A,\  D,\  L_3$&sl(2,R)$\oplus2\textsf{n}_{1,1}$\\
7$^\star\ $&$G(\tilde r)+\kappa\varphi$
\vspace{2mm}
&$L_3+\kappa t,\ G_3, \ P_3$&$\hspace{-2.5mm}\begin{array}{l}\textsf{s}_{5,14}\ \ \text{if}\ \ \kappa\neq0,\\\textsf{n}_{4,1}\oplus\textsf{n}_{1,1}\ \ \text{if}\ \ \kappa=0\end{array}$\\
8$\hspace{2mm} $&
\vspace{2mm}
$\frac1{\tilde
r^{2}}G(\tilde r^\kappa
e^{-\varphi})$
 \vspace{2mm}
&$D+\kappa L_3,\ G_3,\
P_3$ &$\textsf{s}_{5,38}$\\
9$^{\star\ast}$&\vspace{2mm}$\frac1{\tilde r^2}G(\varphi)$&$A,\ D,\   G_3,\  P_3$&sl(2,R)$  {{\subset}\hspace{-3mm} + } \textsf{n}_{3,1}$\\
10$^\star $&$G(x_1)$
\vspace{2mm}
&$G_3,\ P_3,\  P_2,\ G_2,\ L_1$&g(1,2)\\
11$\hspace{2mm} $&
\vspace{2mm}
$G(r)$&$L_1,\  L_2,\  L_3$ &so(3)$\oplus  2\textsf{n}_{1,1}$\\
12$^\ast\ $&\vspace{2mm}$\frac\kappa {r^{2}}$& $A,\ D,\
L_1, \ L_2, \ L_3
$&sl(2,R)$\oplus\ $so(3)$
\oplus\textsf{n}_{1,1} $\\
13$^{\star\ast}$&\vspace{2mm}$\frac\kappa{\tilde r^2}$&$ A,\  D,\   G_3, \ P_3,\ L_3$&sl(2,R)$  {{\subset}\hspace{-3mm} + }  \textsf{n}_{3,1}\oplus\textsf{n}_{1,1}$\\
14$^{\star\ast}$&\vspace{2mm}$\frac\kappa{x_1^2}$&$ A,  D,  G_2, G_3, P_2,  P_3,  L_1 $&schr(1,2)\\

15$\hspace{2mm} $&\vspace{2mm}$\varepsilon\frac{\omega^2x_3^2}2+G(x_1,x_2)$&
$B_3^\varepsilon(\omega),\ \hat B_3^\varepsilon(\omega)\ $&$\hspace{-2.5mm}\begin{array}{l}\textsf{s}_{4,9}\ \text{if}\  \varepsilon=1,\\\textsf{s}_{4,8}\ \text{if}\  \varepsilon=-1\end{array}$\\
16$\hspace{2mm} $&\vspace{0mm}$\varepsilon\frac{\omega^2x_3^2}2+G(\tilde r)+\mu\varphi$&$B_3^\varepsilon(\omega),\ \hat B_3^\varepsilon(\omega),\   L_3+\mu t$&$\hspace{-2.5mm}\begin{array}{l}\textsf{s}_{5,16}\ \text{if}\ \ \varepsilon=1, \mu\neq 0\\\textsf{s}_{5,15}\ \text{if}\  \varepsilon=-1, \mu\neq 0\\\textsf{s}_{4,9}\oplus\textsf{n}_{1,1}\ \text{if}\ \ \varepsilon=1, \mu=0,\\\textsf{s}_{4,8}\oplus\textsf{n}_{1,1}\ \text{if}\  \varepsilon=-1, \mu=0\end{array}$\\
17$^\star\ $&\vspace{2mm}$\varepsilon\frac{\omega^2x_2^2}2+G(x_1)$&$B_2^\varepsilon(\omega),\ \hat B_2^\varepsilon(\omega),\   P_3,\ G_3$&$\hspace{-2.5mm}\begin{array}{l}\textsf{s}_{6,160} \ \text{ if }\ \varepsilon=-1,\\\textsf{s}_{6,161} \ \text{ if }\ \varepsilon=1\end{array}$\\
18$\hspace{2mm} $&\vspace{2mm}$\varepsilon_1\frac{\omega_1^2x_1^2}2+
\varepsilon_2\frac{\omega_2^2x_2^2}2+G(x_3)$&$B_k^{\varepsilon_k}(\omega_k), \hat B_k^{\varepsilon_k}(\omega_k), k=1,2 $&$\hspace{-2.5mm}\begin{array}{l}\textsf{s}_{6,162} \ \text{ if }\ \varepsilon_1=\varepsilon_2=-1,\\\textsf{s}_{6,164} \ \text{ if }\ \varepsilon_1\varepsilon_2=-1, \\\textsf{s}_{6,166} \ \text{ if }\ \varepsilon_1=\varepsilon_2=1\end{array}$\\
19$\hspace{2mm} $&\vspace{2mm}$\varepsilon\frac{\omega^2\tilde r^2}2+G(x_3)$&$B_k^{\varepsilon}(\omega), \hat B_k^{\varepsilon}(\omega), L_3$&$\hspace{-2.5mm}\begin{array}{l}\textsf{s}_{6,m}\oplus\textsf{n}_{1,1}\ \text{with all}\ m \\\text{given in Item 18}\end{array} $\\
20$\hspace{2mm} $&\vspace{2mm}$\varepsilon_1\frac{\omega_1^2x_1^2}2+\varepsilon_2\frac{\omega_2^2x_2^2}2+
\varepsilon_3\frac{\omega_3^2x_3^2}2$&$B_a^{\varepsilon_a}(\omega_a), \hat B_a^{\varepsilon_a}(\omega_a), a=1,2,3 $&$\textsf{s}_{8,1}(\varepsilon_1,\varepsilon_2,\varepsilon_3) $\\
21$^\star\ $&\vspace{2mm}$\varepsilon_1\frac{\omega_1^2x_1^2}2+
\varepsilon_2\frac{\omega_2^2x_2^2}2$&$B_k^{\varepsilon_k}(\omega_k),\ \hat B_k^{\varepsilon_k}(\omega_k),\  P_3,\ G_3  $&$\textsf{s}_{8,2}(\varepsilon_1,\varepsilon_2)$\\
22$\hspace{2mm} $&\vspace{2mm}$\varepsilon\frac{\omega^2\tilde r^2}2+\varepsilon_3\frac{\omega_3^2x_3^2}2$&$ L_3,\ B_a^{\varepsilon_a}(\omega_a),\ \hat B_a^{\varepsilon_a}(\omega_a) $&$\textsf{s}_{9,1}(\varepsilon,\varepsilon_3)$  \\
23$^\star\ $&\vspace{2mm}$\varepsilon\frac{\omega^2\tilde r^2}2$&$ G_3, P_3,   L_3, B_k^{\varepsilon}(\omega), \hat B_k^{\varepsilon}(\omega)
$&$\textsf{s}_{9,2}(\varepsilon)$\\
24$^\star\ $&
\vspace{2mm}$\varepsilon\frac{\omega^2x_3^2}2$&$B_3^\varepsilon(\omega), \hat B_3^\varepsilon(\omega),   P_k, G_k,  L_3 $&$\textsf{s}_{9,3}(\varepsilon)$\\

\hline
\hline
\end{tabular}

All presented systems by construction admit symmetries $P_0$ and $I$, the latter is  the unit operator. Additional symmetries are presented in Columns 3, where $P_a,\ L_a=\frac12\varepsilon_{abc}M^{bc}, \ D,\ A, B^\varepsilon_a(\omega_a)$ and $\hat B^\varepsilon_a(\omega_a)$ are generators presented in (\ref{sok}), (\ref{sos}), (\ref{sos2}) and (\ref{sos3}).
 The related symmetry algebras
are fixed in the fourth columns, where
 $\textsf{n}_{a,b}$ and $\textsf{s}_{a,b}$ are nilpotent and solvable Lie
 algebras of dimension $a$.   To identify these algebras for $a\leq 6$ we use the notations presented in \cite{snob}. The symbol $2\textsf{n}_{1,1}$ (or $3\textsf{n}_{1,1}$) denotes the direct sum of two (or three)  one-dimension algebras. In addition, g(1,2) and shcr(1,2) are Lie algebras of Galilei and Schr\"odinger groups in (1+2) dimensional space.

In the tables we specify also the admissible equivalence transformations additional to ones belonging to the extended Euclid group.  Namely, the star near the item number means that the corresponding Schr\"odinger equation admits additional equivalence transformation (\ref{et3}) for independent variables $x_a$ provided $\frac{\p V}{\p x_a}=0$. The asterix  marks the items which specify equations  admitting transformation (\ref{et1}) and (\ref{et2}).

The algebras of symmetries presented in Items 20--24 are solvable and have dimension $d\geq8$. We denote them formally as $\textsf{s}_{d,a}(.)$ without refereing to any data base,  since the classification of such algebras is still far from the completeness. Let us present  commutation relations which specify these algebras:
\begin{gather}\la{crr}\begin{split}&[P_0,B_a^\varepsilon]=\ri \omega\hat B_a^\varepsilon,\quad [P_0,\hat B_a^\varepsilon]=\ri \varepsilon \omega B_a^\varepsilon, \quad [P_0,G_3]=\ri P_3, \quad [B_a^\varepsilon,\hat B_b^\varepsilon]=\ri\delta_{ab} I,\\&[B_2^\varepsilon,L_3]=\ri B_1^\varepsilon,\quad [B_1^\varepsilon,L_3]=-\ri B_2^\varepsilon,\quad [\hat B_2^\varepsilon,L_3]=\ri \hat B_1^\varepsilon,\quad [\hat B_1^\varepsilon,L_3]=-\ri \hat B_2^\varepsilon\end{split}\end{gather}
where only non-trivial commutators are presented.

Thus we classify all non-equivalent Lie  symmetries which can be admitted by 3$d$ Schr\"odinger equations (\ref{H}). Some of them are new, see discussion in the following section. 

Symmetries of the 2$d$ equation can be obtained making reduction of the results presented in Table 1 with identifying of reduced symmerty algebras. In this way we come to the results presented in Table 2.
\begin{center}Table 2.
Non-equivalent potentials and  symmetries for 2$d$ Schr\"odinger equation.

\begin{tabular}{l l l l }
\hline No&Potential $V$&Symmetries&Invariance algebras
\\
\hline

1$\hspace{2mm} $&$G(\tilde r)+\kappa\varphi$
\vspace{1mm}
&$ L_3+\kappa t
$&$\begin{array}{c}\textsf{n}_{3,1}\ \ \text{if}\ \ \kappa\neq0,\\
3\textsf{n}_{1,1}\ \text{if}\ \ \kappa=0\end{array}$\\
2$\hspace{2mm} $&$\frac1{{\tilde r}^2} G(r^\kappa e^
{-\varphi})$
 \vspace{1mm}
 &$D+\kappa L_3$&$\begin{array}{c}\textsf{s}_{2,1}\oplus\textsf{n}_{1,1}\end{array}$\\
3$^\ast\ $&$\frac1{ \tilde r^2}G(\varphi)$
\vspace{1mm}
&$A,\quad D$& \hspace{2mm}sl(2,R)$\oplus\textsf{n}_{1,1}$
\\
4$^\star\ $&$G(x_1)$
\vspace{1mm}
&$P_2,\ G_2,$&\hspace{2mm}$\textsf{n}_{4,1}$\\
5$^\ast\ $&\vspace{1mm}$\frac\kappa {{\tilde r}^{2}}+\kappa\varphi$& $A,\ D,\
\ L_3+\kappa t
$&$ \begin{array}{c}\text{sl(2,R)}\oplus\ \textsf{n}_{3,1}\ \ \text{if}\ \ \kappa\neq0,\\
\text{sl(2,R)}\oplus\ 3\textsf{n}_{1,1}\ \text{if}\ \ \kappa=0\end{array}$\\
6$^{\star\ast}$&\vspace{1mm}$\frac\kappa{x_1^2}$&$ A,\  D,\  G_2,\  P_2, $&\hspace{2mm}sl(2,R)$\oplus\textsf{n}_{4,1}$\\
7$\hspace{2mm} $&\vspace{1mm}$\varepsilon\frac{\omega^2x_1^2}2+G(x_2)$&$B_1^\varepsilon(\omega),\ \hat B_1^\varepsilon(\omega)\ $&$\begin{array}{l}\textsf{s}_{4,9}\ \text{if}\  \varepsilon=1,\\\textsf{s}_{4,8}\ \text{if}\  \varepsilon=-1\end{array}$\\
8$\hspace{2mm} $&\vspace{1mm}$\varepsilon\frac{\omega^2x_1^2}2$&$B_1^\varepsilon(\omega),\ \hat B_1^\varepsilon(\omega),\ P_2,\ G_2$&$\begin{array}{l}\textsf{s}_{6,160} \ \text{ if }\ \varepsilon=-1,\\\textsf{s}_{6,161} \ \text{ if }\ \varepsilon=1\end{array}$\\
9&\vspace{1mm}$\varepsilon_1\frac{\omega_1^2x_1^2}2+
\varepsilon_2\frac{\omega_2^2x_2^2}2$&$B_1^{\varepsilon_1}(\omega_1),\ B_2^{\varepsilon_2}(\omega_2),\  \hat B_1^{\varepsilon_1}(\omega_1),\ \hat B_2^{\varepsilon_2}(\omega_2)$&$\begin{array}{l}\textsf{s}_{6,162} \ \text{ if }\ \varepsilon=\varepsilon'=-1,\\\textsf{s}_{6,164} \ \text{ if }\ \varepsilon\varepsilon'=-1, \\\textsf{s}_{6,166} \ \text{ if }\ \varepsilon=\varepsilon'=1\end{array}$\\
\hline
\hline
\end{tabular}
\end{center}

\section{Discussion}

Our revision of continuous point symmetries of the main equation of quantum mechanics is seemed  to be useful. We recover the  classical results presented in \cite{Boy}, but also find four systems, missing in the Boyer classification. These systems are represented in Items 1, 2, 7 and 16 of Table 1 and Items 1, 5 of Table 2. 

Thus we present a correct list of inequivalent point continuous  symmetries and the corresponding potentials which can be admitted by one-particle Schr\"odinger equation. This list does not include the infinite symmetry group of transformations $\psi\to\psi+\tilde\psi$ where $\tilde\psi$ is an arbitrary solution of equation (\ref{H}). In accordance with the superposition rule,  such evident symmetries are valid for all linear equations.

In the classification table the systems admitting additional equivalence transformations are clearly indicated. The exact identification of the invariance algebras is presented also with using notations proposed in \cite{snob}. We believe it is an important and useful information about the symmetry accepted systems. In particular, using this identification, the reader interested in the Casimir operators of the symmetry algebras can easy find them in book \cite{snob}.

Notice that the low dimension algebras of dimension $d\leq5$ and
some class of the algebras of dimension 6 had been classified
by
 Mubarakzianov \cite{mur}, see also  more contemporary
 and accessible papers \cite{bas}, \cite{boy1} and \cite{boy2} were his results are slightly  corrected.

A natural question arises: why the new systems presented here were not recognized by Boyer? \footnote{I am indebted to Prof. W. Miller, Jr for asking   me this question .}  For readers familiar with paper \cite{Boy} it is possible to indicate two miss points  there. First, in equation (2.2g) giving the generic form of rotation generators the admissible term $C_{ij} t$ with constant antisymmetric tensor $C_{ij}$ is missing. This term cannot be reduced to zero if parameters $g_i$ present in this formula are trivial. Secondly, Boyer did not use a list of non-equivalent subalgebras of algebra $\tilde{\text e}$(3), and as a result the symmetry presented  in Item 2 of Table 1 was overlooked. 

In any case, Charles  Boyer was the first who made the group classification of 3$d$ Schr\"odinger equations with arbitrary potentials. Moreover, he deduced the determining equations (\ref{ded}) for  arbitrary number of spatial variables. Up to minor misprints, the list of non-equivalent symmetries presented by him is correct but incomplete, and I appreciate the chance to make a small addition to these well known results. For group classification of nonlinear Schr\"odinger equations and their conditional symmetries see papers \cite{pop}, \cite{N1} and \cite{FN}

We restrict our analysis to equations (\ref{H}) with two and three spatial variables. Its extension to equations with more variables is not too difficult but more cumbersome  problem, see the the last paragraph of the Appendix.
\renewcommand{\theequation}{A\arabic{equation}} %
\setcounter{equation}{0}
\appendix
\section{Appendix. Some details of calculations}

We will not reproduce detailed calculations requested to solve the determining equations but  present some points of the algebraic approach which was used to do it.

Considering various differential consequences of equation (\ref{ded}) it is possible to show  that up to equivalence transformations (\ref{et1})--(\ref{et3})  the generic symmetry (\ref{so}) with coefficients (\ref{kil}), (\ref{eta})  is nothing but a linear combination of symmetries (\ref{sok}), (\ref{sos}),  (\ref{sos2}), (\ref{sos3}) and yet indefinite function $f$. Thus to find all non-equivalent solutions of equation (\ref{ded}) we have to go over such (non-equivalent) combinations, which are restricted by the following condition:
 the corresponding functions $\alpha, \ \nu^a,$  $\theta^{ab}$ and $f$ should be proportional to the same function $\Phi(t)$.  In accordance with (\ref{sok}), (\ref{sos}),  (\ref{sos2}) and  (\ref{sos3}) this function can be scalar, linear in $t$, trigonometric or hyperbolic.

 The scalar function $\Phi(t)$ corresponds to  linear combinations of generators $P_a,\ L_a=\frac12\varepsilon_{abc}$ and $D$  presented by equation (\ref{sok}). These generators form a basis of extended   Euclidean algebra $\tilde{\text{e}}$(3), whose non-equivalent subalgebras has been classified in \cite{baran}. In particular, this algebra has four non-equivalent one-dimensional subalgebras spanned on the following generators:
\begin{gather}\la{1d}L_3=M_{12},\quad L_3+P_3,\quad D+\mu L_3,\quad P_3.\end{gather}
The corresponding non-zero coefficients in equation (\ref{ded}) are $\theta^{12}=1$ for $L_3$, $\theta^{12}=\nu^3=1 $ for $L_3+P_3$, $a=1, \ \theta^{12}=\mu$ for $D+\mu L_3$ and $\nu^3=1$ for $P_3$. In addition,  arbitrary function $f$ should be linear in $t$, i.e., $\eta=f=\kappa t$, and just this function can be added to all generators (\ref{1d}). In particular, let $Q=L_3+\kappa t$ then equation (\ref{ded}) is reduced to the following form: \begin{gather*}L_3V=-\ri\kappa\end{gather*}
or
\begin{gather*}\frac{\p V}{\p \varphi}=\kappa,\quad \varphi=\arctan\left(\frac{x_2}{x_1}\right),\end{gather*}
and so $V=\kappa\varphi$. Just this case is missing in Boyer classification.

Solving such defined class of equations (\ref{ded}) we obtain potentials presented in Items 1--5 of Table 1. In Item 4 we set $\kappa=0$ since this parameter  can be reduced to zero using mapping inverse to (\ref{et3}). The additional symmetry $G_3$ is presented there since it generates the same equation for potential as $P_3$. In Item 5 we also have the additional symmetry $A$ which requests the same potential as symmetry $D$. 

In general, it is possible to fix the following pairs  and triplets of "friendly symmetries"
 \begin{gather}\langle P_a, G_a\rangle, \quad \langle A, D\rangle,\quad \langle B^\varepsilon_a, \hat B^\varepsilon_a\rangle\quad \langle (P_1,P_2), L_3 \rangle,\quad \langle ( B^\varepsilon_1, B^\varepsilon_2),L_3\rangle,\quad  \end{gather}
 If equation (\ref{H}) admit a symmetry from one of the presented pairs, it automatically admit also the other symmetry. Symmetries presented in brackets induce the third symmetry from the triplet.

The next step is to exploit the non-equivalent two-dimensional subalgebras of the extended Euclid algebra, which can be spanned on the following basis elements:
\begin{gather}\la{2d} \langle L_3+\kappa t, P_3 \rangle,\quad \langle D+\kappa L_3, P_3 \rangle,\quad \langle P_2, P_3 \rangle,\quad \langle D, L_3 \rangle.\quad\end{gather}

 Since any sets (\ref{2d}) includes at least one  basis element from list (\ref{1d}), we have to solve equation (\ref{ded}) generated by the second element for potentials presented in Items 1--5 of Table 1. As a result we obtain potentials included in Items 6--10. The corresponding symmetry algebras are  extended at the cost of "friendly symmetries".

Analogously, considering non-equivalent three dimensional subalgebras
\begin{gather}\la{3dim} \langle D, P_3, L_3\rangle,\quad  \langle D, P_1,  P_2\rangle, \quad \langle L_1, L_2, L_3\rangle,\quad \langle L_3, P_1,  P_2\rangle \\\langle P_1, P_2,  P_3\rangle,\quad \langle L_3+ P_3,\ P_1,\ P_2\rangle,\quad \langle D+\mu L_3, P_1,  P_2\rangle,\ \mu>0\la{empty}\end{gather}
we obtain potentials presented in Items 11--14 and recover once more the cases given in Items 7, 8, 10. Notice  that now we have also a simple algebra so(3) realized by $L_1, L_2$ and $L_3$. In addition,  algebras (\ref{empty}) are valid only for scalar potentials.

Thus we have described all symmetry algebras including generators with time independent coefficients $\xi^a$. In fact some of them include the coefficients linear in $t$, but such generators appear automatically as "friendly symmetries". And if we now will consider specially realizations linear in $t$, the list of potentials presented in Items 1--14 will not be extended.

To obtain the remaining part of the table we have to search for symmetries including hyperbolic functions. This job is reduced to simple enumeration of possibilities with one, two, or three pairs of operators $B^\varepsilon_a, \hat B^\varepsilon_a$ with the same or different frequency parameters $\omega=\omega_a$.

Let us note that our analysis can be directly extended to the case of equations (\ref{H}) with more large number $n$ of spatial variables.  For example, for $n=4$ we have to start with the following one-dimensional subalgebras:
\begin{gather*} P_4, \ M_{12}, \ M_{12}+M_{34},\  M_{12}+\alpha M_{34} \ (0<\alpha<1),\ M_{12}+\nu P_4 \ (\nu>0), \\ D+\lambda M_{12}, \ D+M_{12}+M_{34},\ M_{12}+\alpha M_{34} +\beta D \ (0<\alpha\leq1,\ \beta>1)\end{gather*} whose number is more extended than in $3d$ case (compare  equation (\ref{1d}) but not dramatically large. On the other hand,  the determining equation (\ref{ded}) is defined for arbitrary $n$.

\end{document}